\documentclass{article}

\usepackage{amsmath}
\usepackage{amsthm}
\usepackage{amssymb}
\usepackage{graphicx}
\usepackage{psfrag}
\usepackage{framed}

\newcommand{\aut}{\operatorname{Aut}}
\newcommand{\poly}{\operatorname{poly}}

\newcommand{\uinvnorm}{|\kern-2pt|\kern-2pt|}

\theoremstyle{plain}
\newtheorem{theorem}{Theorem}[section]

\newtheorem{lemma}[theorem]{Lemma}

\newtheorem{corollary}[theorem]{Corollary}

\theoremstyle{definition}
\newtheorem{definition}[theorem]{Definition}
\newtheorem{example}[theorem]{Example}

\theoremstyle{remark}
\newtheorem{remark}[theorem]{Remark}

\begin{document}

\title{{\sf\bfseries Quantum Algorithms and Covering Spaces}}

\author{{\sf Tobias J.\ Osborne\footnote{\texttt{T.J.Osborne@bristol.ac.uk}}}\\ {\sf School of Mathematics}\\ {\sf University of Bristol}\\ {\sf University Walk}\\ {\sf Bristol BS8 1TW}\\ {\sf United
Kingdom} \and {\sf Simone
Severini\footnote{\texttt{ss54@york.ac.uk}. Present address: Department of Mathematics and Department of Computer Science, University of York, York YO10 5DD, UK}}\\
{\sf
Department of Computer Science}\\ {\sf University of Bristol}\\
{\sf Merchant Venturers' Building}\\ {\sf Bristol BS8 1UB}\\ {\sf
United Kingdom}}
\date{{\sf \today}}
\maketitle

\begin{abstract}
It's been recently demonstrated that quantum walks on graphs can
solve certain computational problems faster than any classical
algorithm. Therefore it is desirable to quantify those purely
combinatorial properties of graphs which quantum walks take
advantage of and try and separate them from those properties due
to the encoding of the problem. In this paper we isolate the
combinatorial property responsible (at least in part) for the
computational speedups recently observed. We find that
continuous-time quantum walks can exploit the \emph{covering
space} property of certain graphs. We formalise the notion of
graph covering spaces. Then we demonstrate that a quantum walk on
a graph $Y$ which covers a smaller graph $X$ can be equivalent to
a quantum walk on the smaller graph $X$. This equivalence occurs
only when the walk begins on certain initial states,
\emph{fibre-constant states}, which respect the graph covering
space structure. We illustrate these observations with walks on
Cayley graphs; we show that walks on fibre-constant initial states
for Cayley graphs are equivalent to walks on the induced Schreier
graph. We also consider the problem of constructing efficient gate
sequences simulating the time evolution of a continuous-time
quantum walk. We argue that if
$Y\xrightarrow{\pi_{N}}X_{N}\xrightarrow{\pi_{N-1}}X_{N-1}
\xrightarrow{\pi_{N-2}}\cdots\xrightarrow{\pi_{1}}X_1$ is a tower
of graph covering spaces satisfying certain uniformity and growth
conditions then there exists an efficient quantum gate sequence
simulating the walk. For the case of the walk on the $m$-torus
graph $T^m$ on $2^n$ vertices we construct a gate sequence which
uses $O(\poly(n))$ gates which is independent of the time $t$ the
walk is simulated for (and so the sequence can simulate the walk
for \emph{exponential} times). We argue that there exists a wide
class of nontrivial operators based on quantum walks on graphs
which can be measured efficiently using phase estimation.
Interestingly, measuring these operators won't be unitarily
equivalent to the quantum fourier transform. Finally, motivated by
our results we introduce a new general class of computational
problems, {\sf HiddenCover}, which includes a variant of the
general hidden subgroup problem as a subclass. We argue that
quantum computers ought to be able to utilise covering space
structures to efficiently solve problems from {\sf HiddenCover}.
\end{abstract}

\section{Introduction}
There is a growing belief that quantum computers can solve certain
computational problems exponentially faster than any classical
computer. Strong evidence for this belief comes in the form of
Shor's algorithm \cite{shor:1994a} for factorisation, and the
graph traversal algorithm of Childs \emph{et.\ al.}\
\cite{childs:2003a}.

Despite the spectacular success of the known quantum algorithms we
believe that there is still only a fairly rudimentary
understanding of the properties of quantum mechanics which are
useful for computational speedups. We ascribe this poor
understanding to at least two causes: (i) we are unsure what sorts
of problems might be amenable to quantum-computational speed up;
and (ii) even if we firmly believed that an efficient quantum
algorithm existed for a problem, it is hard to come up with this
putative algorithm because it is difficult (at least for us) to
reason within the traditional quantum computing model --- the
quantum circuit model.

All quantum algorithms are traditionally expressed in the quantum
circuit model (see \cite{nielsen:2000a} and \cite{preskillnotes}
for a detailed description of the quantum circuit model). The
quantum circuit model assigns a unit cost to certain elementary
quantum gates, such as, for example, {\sc cnot}, $H$, and $T$ (the
$\pi/8$ phase gate). This is by no means the only way to express
quantum algorithms. Three other computing paradigms are
polynomially equivalent to the quantum circuit model: the quantum
Turing machine model \cite{bernstein:1997a}, the one-way quantum
computer \cite{briegel:2001b}, and the adiabatic evolution model
(which has recently been shown to be polynomially equivalent to
the quantum circuit model in \cite{aharonov:2004a}).

Why might one want to consider computational paradigms other than
the quantum circuit model? The reason is that there are conceptual
peculiarities with the quantum circuit model which make it hard to
work with when designing algorithms the traditional way. Firstly,
there is a back-action that quantum superposition induces in gates
like the {\sc cnot} where there can be a ``backwards'' flow of
quantum information for certain initial states. Secondly, the
existence of quantum entanglement seriously confuses the causal
structure of quantum circuits because we can think of a Bell pair
as sending a qubit backwards in time! However, in favour of the
quantum circuit model is the fact that exponentially many degrees
of freedom can be summarised succinctly.

Some of the negative features of the quantum circuit model are
amply addressed in the quantum walk model of computation. Quantum
walks, or the quantum dynamics of a free particle hopping on a
graph, are an exciting new paradigm for quantum computing. (For a
review of quantum walks see \cite{kempe:2003b} and references
therein.) The attractiveness of quantum walks is that they provide
an extremely intuitive way to \emph{visualise} exponentially many
quantum degrees of freedom. In contrast to the quantum circuit
model, in a quantum walk there is a clear physical picture of
\emph{where} information is flowing and a definite notion of
\emph{cause and effect}. The price we pay for these intuitive
features is that the exponentially ($2^n$ for $n$ qubits) many
degrees of freedom of $n$ qubits translate to exponentially
vertices in the graph. Additionally, there is no clear way to
translate a quantum walk efficiently into the quantum circuit
model. Despite these difficulties we believe that quantum walks
provide an attractive methodology in the quantum algorithm
designer's toolkit because they appeal to the geometrical
intuitions.

The growing number of quantum-walk algorithms might be seen as
evidence for the conceptual utility of the quantum walk model. We
mention three recent algorithms: (i) the quantum walk search
algorithm \cite{shenvi:2003a}, \cite{childs:2003c}, and
\cite{ambainis:2004a}; (ii) the graph-traversal algorithm of
Childs \emph{et.\ al.}\ \cite{childs:2003a}; and (iii) the element
distinctness algorithm of Ambainis \cite{ambainis:2003a}, and
relatives \cite{childs:2003b, magniez:2003a, szegedy:2004b}.

What unites the quantum walk algorithms? (And, more ambitiously,
all quantum algorithms?) Let's concentrate on the results of
\cite{childs:2002a, childs:2003a, moore:2002a, kempe:2003a}, which
are based on the continuous-time quantum walk. (Note that the
quantum walks on the hypercube \cite{moore:2002a, kempe:2003a} and
the graphs in \cite{childs:2002a} don't provide the speedups for
the solution to any algorithmic problem.) The results in these
papers appear to be related phenomena (they all take advantage of
``column spaces''), however, this relationship has not yet, to the
best of our knowledge, been quantified fully. In this paper we
identify and generalise the combinatorial property of these graphs
which leads to small hitting times. We believe this is a key
ingredient underlying quantum speedups of hitting times. The
combinatorial property we isolate is that all of the graphs walked
on in these papers are covering spaces for much smaller graphs.

\subsection*{Quantum computers and covering spaces}

\begin{figure*}
\begin{center}
\psfrag{pi}{$\pi$} %
\includegraphics{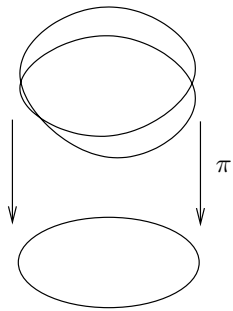}
\caption{An example of a covering space. In this case the circle
$X=S^1$ is covered by another circle $Y$ which has twice the
circumference of $X$ under the projection $\pi$. Note that the
inverse projection of any point $x\in X$ is a finite set
consisting of two points $a$ and $b$ in the covering space $Y$.
One can think of this covering space as a subset of the Riemann
surface for $f(z)=\sqrt{z}$, which is a covering space for
$\mathbb{C}$.}\label{fig:cover}
\end{center}
\end{figure*}

What is a covering space? Suppose that $X$ and $Y$ are
arcwise-connected and locally arcwise-connected topological
spaces, respectively. Then $(Y,\pi)$ is said to be a covering
space of $X$ if $\pi:Y\rightarrow X$ is a surjective continuous
map with every $x\in X$ having an open neighborhood $U$ such that
every connected component of $\pi^{-1}(U)$ is mapped
homeomorphically onto $U$ by $\pi$. The preimage of a point in $X$
is called a \emph{fibre} of $\pi$. An example covering space is
shown in figure~\ref{fig:cover}.

In this paper we will show that a continuous-time quantum walk on
a graph $Y$ which is a covering space for another graph $X$ (in
the natural topology) is isomorphic to a quantum walk on $X$,
under certain specific circumstances. This equivalence occurs when
the walk starts on \emph{fibre-constant} quantum states.
Additionally, we will argue, and show in some cases, that if there
is a tower of covering spaces
\begin{equation}
Y\xrightarrow{\pi_{N}}X_{N}\xrightarrow{\pi_{N-1}}X_{N-1}
\xrightarrow{\pi_{N-2}}\cdots\xrightarrow{\pi_{1}}X_1
\end{equation}
then there is an efficient gate sequence (i.e.\ using
$O(\poly(n))$ elementary gates) for the quantisation of $Y$. The
idea we exploit is to recursively and hierarchically construct the
gate sequence from the ``elementary gate'' $U(X_1)$ and the
specification of how $U(X_1)$ lives in $U(X_2)$.

We believe that these results are not specific to continuous-time
quantum walks, but rather \emph{indicative} of a general
principle.

We formulate this principle in the following way: consider some
collection of mathematical objects like the class of simple
graphs, or something with more structure, like the class of finite
fields. Let $U(X)$ be a quantisation scheme for this class of
objects, by which we mean a way to associate a unitary matrix
$U(X)$ acting on a finite-dimensional Hilbert space
$\mathcal{H}(X)$ with any object $X$. Suppose, further, that $Y$
is a covering space $\pi:Y\rightarrow X$ for another object $X$.
If the quantisation scheme $U(X)$ is ``sufficiently well-behaved''
then $U(Y)$ should be related to $U(X)$ according to a map, the
\emph{pull-back} $\pi^*:U(X)\rightarrow U(Y)$. If, further, $Y$ is
determined from $X$ in a sufficiently simple way, then it should
be possible to construct $U(Y)$ from $U(X)$ according to this
specification. Mimicking the recursive and hierarchical
construction for gate-sequences for continuous-time quantum walks
we expect that if
\begin{equation}
Y\xrightarrow{\pi_{N}}X_{N}\xrightarrow{\pi_{N-1}}X_{N-1}
\xrightarrow{\pi_{N-2}}\cdots\xrightarrow{\pi_{1}}X_1
\end{equation}
is a tower of covering spaces obeying some uniformity and growth
conditions then there should be an efficient gate sequence for the
quantisation $U(X)$.

We retroactively argue that there is an example of where this
philosophy has already proven successful. In this example we take
as our space of mathematical objects the category
$\mathbf{GrpFin}$ of finite groups. We take the quantisation
scheme to be the contravariant functor which associates the
quantum fourier transform with every finite group. (For an
introduction to category theory see \cite{maclane:1998a}.) For
certain towers of subgroups, i.e.\ those which are
\emph{polynomially uniform} \cite{moore:2003a}, there is an
efficient hierarchical scheme to construct the quantisation
$U(Y)$. Here we are taking the ``is a subgroup'' relation to be
the covering space relation in this category. To accord this
exactly with the definition of covering space given earlier
requires the introduction of an appropriate topology on a finite
group. This is something we will avoid, preferring only this
qualitative argument  --- clearly the analogy isn't perfect.

What sort of problems could these supposed efficient quantisations
solve? We will argue that they could be used to solve a problem we
call the \emph{Hidden Covering Space} problem which includes the
hidden subgroup problem as a subclass. Roughly speaking, if there
is a function $f$ on $Y$ which is periodic on some object $X$
covered by $Y$, then a quantum computer should be able to identify
the structure $X$ efficiently.

The general properties of covering spaces are extremely
tantalising and suggestive. We hope to convince the reader that
the results we have found concerning graph covering spaces and
continuous-time quantum walks are indicative of a much larger
framework. We believe that many quantum algorithms exhibiting an
exponential separation between classical and quantum complexity
could exist. We argue that the properties of quantisations of
covering spaces may provide many opportunities to design such new
algorithms. This is not least because covering spaces are
pervasive in mathematics, from number fields and algebraic
surfaces to geometric group theory, and, of course, Galois theory.

The outline of this paper is as follows. We begin in
\S\ref{sec:graphquant} by describing the quantisation scheme we
study in the remainder of this paper, the continuous-time quantum
walk. In \S\ref{sec:graphcover} we review the theory of graph
covering spaces and the heat kernel for graphs. We then apply
these results in \S\ref{sec:indqw} to demonstrate that quantum
walks which begin on fibre-constant states are isomorphic to
quantum walks on smaller graphs. We illustrate our results in
\S\ref{sec:examples} for the hypercube and Cayley graphs. In
\S\ref{sec:efficient} we utilise the covering-space properties of
the $m$-torus graph on $2^n$ vertices to construct an efficient
gate sequence for their quantisation. Motivated by our results we
introduce, in \S\ref{sec:hiddencover}, a new class of problem,
{\sf HiddenCover}, which quantum computers may be able to solve
efficiently. We also solve the hidden cover problem for the
$m$-torus graph which, incidentally, provides a (philosophically)
different way to solve the abelian hidden subgroup problem.

\section{Quantisations of Graphs: the Continuous-Time Quantum
Walk}\label{sec:graphquant}

In this section we introduce the \emph{continuous-time quantum
walk}, which is a quantisation scheme for the class of simple
graphs. For further details about the graph-theoretic notation and
terminology we use in this section and the rest of this paper see
\cite{biggs:1993a, cvetkovic:1995a, chung:1997a}.

Let us begin by defining the main objects of our study. By a
\emph{weighted graph} $Y$ we mean a \emph{vertex set} $V=V(Y)$
with an associated \emph{weight function} $w: V\times
V\rightarrow\mathbb{R}^+$ ($\mathbb{R}^+$ denotes the real numbers
$x\in\mathbb{R}$ such that $x\ge0$) which satisfies
\begin{equation}
w(u,v) = w(v,u).
\end{equation}
If $w(u,v)>0$ then we refer to $\{u,v\}$ as an \emph{edge} of $Y$,
and we say that $u$ and $v$ are \emph{adjacent}. By a \emph{simple
graph} we mean the special situation where $w(u,v)$ is either $0$
or $1$ and $w(u,u)=0$ for all $u\in V$.

We define the \emph{degree} $d_v$ for a vertex $v$ to be
\begin{equation}
d_v = \sum_{v\in V} w(u,v).
\end{equation}
A graph is \emph{regular} if all the degrees are the same.

The \emph{Laplacian} of a weighted graph $Y$ on $n$ vertices and
weight function $w$ is the $n\times n$ matrix $\triangle$ given by
\begin{equation}
\triangle_{u,v}\triangleq\begin{cases}d_v - w(v,v)\quad & \text{if
$u=v$},\\-w(u,v)\quad &\text{if $u$ and $v$ are adjacent,} \\
0\quad &\text{otherwise,}
\end{cases}
\end{equation}
where $u$ and $v$ are two arbitrary vertices in $V$.

We now consider introduce the (formal) complex vector space
$\mathcal{H}(Y)$ spanned by the vectors $|u\rangle$, $u\in V$,
which we take to be orthonormal under the natural inner product:
$\langle u|v \rangle = \delta_{u,v}$. Of course this vector space
is isomorphic to $\mathbb{C}^n$. We think of the vector space
$\mathcal{H}(Y)$ as the space of all complex-valued functions $f$
on the finite set $V(Y)$, where for each vector $|f\rangle =
\sum_{u\in V(Y)}f_u |u\rangle \in \mathcal{H}(Y)$ we define the
function $f:V(Y)\rightarrow \mathbb{C}$ by $f(u) = \langle u |
f\rangle$.

We define the \emph{adjacency matrix} of a weighted graph $Y$ to
be the operator
\begin{equation}
A(Y) \triangleq \sum_{u,v \in V(Y)} w(u,v) |u\rangle \langle v|.
\end{equation}

The Laplacian, acts in a natural way on $\mathcal{H}(Y)$ as
\begin{equation}
\triangle = \sum_{u\in V(Y)} (d_u-w(u,u)) |u\rangle\langle u| -
\sum_{\substack{u,v\in V(Y)\\ u\sim v}} w(u,v) |u\rangle\langle
v|,
\end{equation}
where we use the notation $u\sim v$ to mean that $u$ is adjacent
to $v$ and $u\not= v$. For a specific vector $|f\rangle
=\sum_{u\in V(Y)}f_u |u\rangle$ we have
\begin{equation}
\triangle|f\rangle = \sum_{\substack{u,v\in V(Y) \\ u\sim v}}
(f(u)-f(v))w(u,v)|u\rangle.
\end{equation}
It is worth noting that the Laplacian can be written $\triangle =
D(Y) - A(Y)$, where $D(Y) \triangleq \sum_{u,v \in V(Y)} w(u,v)
|u\rangle \langle u|$.

We define the \emph{heat equation} and \emph{Schr\"odinger
equation} for $Y$ to be the differential equations
\begin{equation}\label{eq:heat}
\frac{\partial}{\partial \tau} |\psi(\tau)\rangle =
-\triangle|\psi(\tau)\rangle,
\end{equation}
and
\begin{equation}\label{eq:schroe}
i\frac{\partial}{\partial t} |\psi(t)\rangle =
\triangle|\psi(t)\rangle,
\end{equation}
respectively. Note that the Schr\"odinger equation is equivalent
to the heat equation with the replacement $\tau = it$.

The Laplacian $\triangle$ for a graph on $n$ vertices is a
symmetric matrix and so we can write its spectral decomposition
\begin{equation}
\triangle = \sum_{j=0}^{n-1} \lambda_j |\lambda_j\rangle \langle
\lambda_j|.
\end{equation}
We often refer to an eigenstate of $\triangle$ as a \emph{harmonic
eigenfunction}.

The spectral decomposition of the Laplacian allows us to define
the \emph{heat kernel} and \emph{propagator}
\begin{equation}
H(\tau) = \sum_{j=0}^{n-1} e^{-\lambda_j\tau} |\lambda_j\rangle
\langle \lambda_j|,
\end{equation}
and
\begin{equation}
U(t) = \sum_{j=0}^{n-1} e^{-i\lambda_jt} |\lambda_j\rangle \langle
\lambda_j|,
\end{equation}
respectively. Note that $U(t) = H(i\tau)$. Using the the heat
kernel and propagator we can solve the heat and Schr\"odinger
equations (\ref{eq:heat}) and (\ref{eq:schroe}) with initial state
$|\psi(0)\rangle$ by defining $|\psi(\tau)\rangle =
H(\tau)|\psi(0)\rangle$ and $|\psi(t)\rangle =
U(t)|\psi(0)\rangle$:
\begin{equation}
\begin{split}
\frac{\partial}{\partial \tau} |\psi(\tau)\rangle &=
\frac{\partial H(\tau)}{\partial \tau}
|\psi(0)\rangle\\
&= -\triangle H(\tau)|\psi(0)\rangle.
\end{split}
\end{equation}
The result for the propagator follows by substituting $\tau = it$.

\begin{definition}
A \emph{continuous-time quantum walk} on a weighted graph $Y$ is
the propagator $U[Y](t)$ associated with the Laplacian
$\triangle[Y]$ for $Y$.
\end{definition}

\section{Graph Covering Spaces and the Heat Kernel}\label{sec:graphcover}
In this section we review the theory of graph covering spaces and
the heat kernel for weighted graphs. For an introduction to graph
covering spaces see \cite{biggs:1993a} and \cite{godsil:1980a}.
For further results in this area see \cite{chung:1999a} and
\cite{stark:1996a, stark:1999a, terras:1999a, stark:2000a,
terras:2002a}. We follow \cite{chung:1999a} closely for most of
this section. The principle result of this section is that the
spectrum of a graph $X$ covered by another graph $Y$ is contained
in the spectrum of $Y$. The eigenvectors of $X$ also induce
eigenvectors in $Y$.

We begin with some definitions.
\begin{definition}\label{def:graphcover}
Suppose we have two graphs $X$ and $Y$ with weight functions
$w_X(x,y)$ and $w_Y(u,v)$, respectively. We say $Y$ is a
\emph{covering space}\footnote{Note that our definition of
(unramified) graph covering spaces accords with the standard
definition given in the introduction when we take for the topology
of the graphs $X$ and $Y$ the \emph{weak topology}. See
\cite{hatcher:2002a} and \cite{stark:1999a, stark:2000a} for
further discussion and details.} for $X$ (alternatively, $X$ is
\emph{covered} by $Y$) if there is a set map $\pi:V(Y)\rightarrow
V(X)$ satisfying the following two properties:
\begin{enumerate}
\item There is a $\mu\in \mathbb{R}^+$ (the positive real numbers
including $0$), called the \emph{index} of $\pi$, such that for
$u$, $v\in V(X)$ the equality holds:
\begin{equation}
\sum_{\substack{x\in\pi^{-1}(u)\\
y\in\pi^{-1}(v)}} w_Y(x,y) = \mu \sqrt{|\pi^{-1}(u)||\pi^{-1}(v)|}
w_X(u,v).
\end{equation}
\item For $x$, $y\in V(Y)$ with $\pi(x)=\pi(y)$ and $v\in V(X)$,
we have
\begin{equation}
\sum_{z\in\pi^{-1}(v)} w_Y(z,x) = \sum_{z'\in\pi^{-1}(v)}
w_Y(z',y).
\end{equation}
\end{enumerate}
\end{definition}

We now translate Definition~\ref{def:graphcover} into a more
useful statement about the adjacency matrices of $Y$ and $X$. In
doing so we make contact with the theory of equitable partitions
\cite{biggs:1993a}.

We begin by considering an arbitrary projection operator
$\pi:V(Y)\rightarrow V(X)$ and define the corresponding
\emph{pull-back} operator $P:\mathcal{H}(Y)\rightarrow
\mathcal{H}(X)$,
\begin{equation}\label{eq:projproj}
P = \sum_{\substack{u \in V(X) \\ x\in \pi^{-1}(u)}}
\frac{1}{\sqrt{|\pi^{-1}(u)|}} |u\rangle \langle x|.
\end{equation}
(It is easy to establish the identity $(P^\dag P)^2=P^\dag P$ via
multiplication, which shows that $P^\dag P$ is a projector.
Similarly, one can show that $PP^\dag = I_{\mathcal{H}(X)}$. This
implies that $P^\dag$ is an isometric imbedding
$P^\dag:\mathcal{H}(X)\rightarrow\mathcal{H}(Y)$.) What are the
conditions that $P$ must satisfy in order that $\pi$ be a legal
covering map? We answer this question in two steps, by translating
parts (i) and (ii) of Definition~\ref{def:graphcover} into
conditions on $P$.

Using the operator $P$ we define the \emph{quotient matrix}
$A_\pi(X) \triangleq P A(Y) P^\dag$, which we want to identify
with the adjacency matrix of $X$. Expanding this expression and
utilising Definition~\ref{def:graphcover} we obtain
\begin{equation}
\begin{split}
A_\pi(X) &= \sum_{\substack{u \in V(X) \\ l\in
\pi^{-1}(u)}}\sum_{\substack{v \in V(X) \\ m\in
\pi^{-1}(v)}}\sum_{x,y \in V(Y)}
\frac{w_Y(x,y)}{\sqrt{|\pi^{-1}(u)||\pi^{-1}(v)|}} |u\rangle
\langle
l|x\rangle \langle y|m\rangle \langle v| \\
&= \sum_{\substack{u \in V(X) \\ l\in
\pi^{-1}(u)}}\sum_{\substack{v \in V(X) \\ m\in \pi^{-1}(v)}}
\frac{w_Y(l,m)}{\sqrt{|\pi^{-1}(u)||\pi^{-1}(v)|}}|u\rangle\langle
v| \\
&=\sum_{\substack{u \in V(X)}}\sum_{\substack{v \in V(X)}} \mu
w_X(u,v)|u\rangle\langle v|=\mu A(X).
\end{split}
\end{equation}
This expression shows us that the adjacency matrix $w_Y(u,v)$ for
$Y$ satisfies the condition (i) of Definition~\ref{def:graphcover}
to be a covering space for the graph $X$ with adjacency matrix
$w_X(u,v)$. Note that part (i) of Definition~\ref{def:graphcover}
is really saying that $Y$ is a covering space for any graph which
has as its adjacency matrix a scalar multiple of $PA(Y)P^\dag$.

In order to establish the matrix version of part (ii) of
Definition~\ref{def:graphcover} we need to make use of the
following result of Godsil and McKay \cite{godsil:1980a}, which is
the fundamental connection between $P$ and graph covering spaces.
We include a proof for completeness.
\begin{lemma}[Godsil and McKay \cite{godsil:1980a}]\label{lem:godsil1}
The graph $Y$ is a covering space for $X$ if and only if
$PA(Y)=A(X)P$.
\end{lemma}
\begin{proof}
For a vertex $v\in V(Y)$, $1\le j\le |V(X)|$ define $h_{vj}$ to be
the sum of the weights of the edges connecting vertices in
$\pi^{-1}(u_j)$ with $v$, where $u_j\in V(X)$. That is, $h_{vj} =
\sum_{x\in\pi^{-1}(u_j)} w_1(v,x)$. For $1\le j\le |V(X)|$, $v\in
V(Y)$, we have
\begin{equation}\label{eq:pay}
\langle u_j|PA(Y)|v\rangle =
\frac{1}{\sqrt{\pi^{-1}(u_j)}}\sum_{x\in \pi^{-1}(u_j)} w_1(x,v) =
\frac{h_{vj}}{\sqrt{\pi^{-1}(u_j)}}
\end{equation}
and
\begin{equation}\label{eq:axp}
\langle u_j|A(X)P|v\rangle =
\frac{w_2(u_j,u_k)}{\sqrt{\pi^{-1}(u_k)}},
\end{equation}
where $v\in\pi^{-1}(u_k)$.

By comparing (\ref{eq:pay}) and (\ref{eq:axp}) we note that if
$PA(Y) = A(X)P$, then $h_{vj}$ equals
$\sqrt{\frac{|\pi^{-1}(u_j)|}{|\pi^{-1}(u_k)|}}w_2(u_j,u_k)$ for
all $v\in \pi^{-1}(u_j)$ and so $\pi$ satisfies part (ii) of
definition~\ref{def:graphcover}.

For the converse, suppose $\pi$ satisfies the conditions of
Definition~\ref{def:graphcover}. We see that
\begin{equation}
PA(Y) = PA(Y)P^\dag P = A(X) P.
\end{equation}
\end{proof}

It is well known that the spectrum of a graph $Y$ which is a
covering space for $X$ is determined, in part, by the spectrum of
$X$. Let's make this a bit more precise. Suppose we have a
harmonic eigenfunction of $|f\rangle$ of $X$ with eigenvalue
$\lambda$. We can \emph{lift} or \emph{pull back} this
eigenfunction to a harmonic eigenfunction $|g\rangle$ of $Y$, by
defining for each vertex $l\in Y$, $\langle l|g\rangle =
\langle\pi(l)|f\rangle/\sqrt{\pi^{-1}(l)}$. Equivalently,
$|g\rangle = P^\dag|f\rangle$.

We use the notation $|\pi^{-1}(u)\rangle \triangleq
1/\sqrt{|\pi^{-1}(u)|}\sum_{l\in\pi^{-1}(u)} |l\rangle$ and write
\begin{equation}
|g\rangle = \sum_{l\in V(X)} f(l) |\pi^{-1}(l)\rangle.
\end{equation}
It follows from Definition~\ref{def:graphcover} that $|g\rangle$
is a harmonic eigenfunction for $Y$.

\begin{lemma}\label{lem:eveclift}
Let $Y$ be a covering space for $X$ with projection map $\pi$.
Then for any eigenvector $|f\rangle$ and scalar $\lambda$,
$A(X)|f\rangle=\lambda|f\rangle$ if and only if
$A(Y)P^\dag|f\rangle = \lambda P^\dag|f\rangle$.
\end{lemma}
\begin{proof}
If $A(X)|f\rangle = \lambda|f\rangle$, then $P^\dag A(X)|f\rangle
= \lambda P^\dag|f\rangle$ and so $A(Y)P^\dag|f\rangle = \lambda
P^\dag|f\rangle$, by lemma~\ref{lem:godsil1}. If
$A(Y)P^\dag|f\rangle = \lambda P^\dag|f\rangle$, then
$PA(Y)P^\dag|f\rangle=\lambda PP^\dag|f\rangle$ and so
$A(X)|f\rangle = \lambda|f\rangle$.
\end{proof}

We define the Laplacian $\triangle(X)$ for the covered graph $X$
via $\triangle(X) \triangleq PD(Y)P^\dag - PA(Y)P^\dag.$

The main result for this section is the following lemma relating
the heat kernels and propagators of a graph $X$ and an arbitrary
covering space $Y$.
\begin{lemma}\label{lem:heatcover}
Suppose $Y$ is a covering space for $X$. Let $H[Y](\tau)$ and
$H[X](\tau)$ and $U[Y](t)$ and $U[X](\tau)$ denote the heat
kernels and propagators for $Y$ and $X$, respectively. Then we
have
\begin{equation}
H[X](\tau) = PH[Y](\tau)P^\dag, \quad \text{and} \quad U[X](t) =
PU[Y](t)P^\dag.
\end{equation}
\end{lemma}
\begin{proof}
We establish the lemma for the heat kernel. The result for the
propagator follows from substituting $\tau = it$.

Note first that $A(X)^r=PA(Y)^rP^\dag$, which follows from a
simple induction using lemma~\ref{lem:godsil1}.

Expanding the heat kernel in an absolutely convergent power series
in $\tau$ gives:
\begin{equation}
\begin{split}
H[X](\tau) &= \sum_{j = 0}^\infty \frac{(-\tau PA(Y)P^\dag)^j}{j!}\\
&= \sum_{j = 0}^\infty P\frac{(-\tau A(Y))^j}{j!}P^\dag =
PH[Y](\tau)P^\dag.
\end{split}
\end{equation}
\end{proof}

Because we are living in finite Hilbert spaces, where everything
is well-behaved, we don't have to worry about the convergence of
series. For this reason, knowing the heat kernel $H(\tau)$ is
equivalent to knowing the propagator $U(t)$. We take advantage of
this fact by only stating all our results about the propagator in
terms of the heat kernel in \emph{imaginary time}.

\section{Induced Quantum Walks --- the Quotient Walk}\label{sec:indqw}
In this section we show that if a graph $Y$ is a covering space
for another graph $X$ with projection $\pi:V(Y)\rightarrow V(X)$
then a quantum walk on $Y$ which begins on a state \emph{constant}
on fibres of $\pi$ is isomorphic to a quantum walk on $X$. This
result provides an insight into the hitting-time speedups observed
by Kempe \cite{kempe:2003a} and Childs \emph{et.\ al.}\
\cite{childs:2002a, childs:2003a}.

\begin{definition}
Let $Y$ be a covering space for $X$ with projection
$\pi:V(Y)\rightarrow V(X)$. We say that a quantum state
$|\psi\rangle \in \mathcal{H}(Y)$ is \emph{fibre-constant} for
$\pi$ if $|\psi\rangle$ can be expressed as
\begin{equation}
|\psi\rangle = \sum_{u\in V(X)} c_{u} |\pi^{-1}(u)\rangle,
\end{equation}
where we are using the notation
\begin{equation}
|\pi^{-1}(u)\rangle = P^\dag|u\rangle=
\frac{1}{\sqrt{|\pi^{-1}(u)|}}\sum_{x\in\pi^{-1}(u)}|x\rangle
\end{equation}
of the previous section. We write $|\psi\rangle =
P^\dag|\phi\rangle$, where $|\phi\rangle = \sum_{u\in V(X)}
c_u|u\rangle$.
\end{definition}

What is the time evolution of a fibre-constant state
$|\psi\rangle$ for covering map $\pi$? The following simple lemma
shows us that it is isomorphic to a walk on $X$.
\begin{lemma}\label{lem:coverwalk}
Suppose $Y$ is a covering space for $X$ with covering map $\pi$.
Let $|\psi\rangle = P^\dag|\phi\rangle$ be a fibre-constant state.
Then the time evolution of $|\psi(t)\rangle$ on $Y$ is isomorphic
to the time evolution of $|\phi\rangle$ on $X$.
\end{lemma}
\begin{proof}
Let $H[Y](\tau)$ be the heat kernel for $Y$. The time evolution of
$|\psi(\tau)\rangle$ on $Y$ is given by
\begin{equation}
\begin{split}
|\psi(\tau)\rangle &= H[Y](\tau) P^\dag|\phi\rangle \\
&= P^\dag H[X](\tau)|\phi\rangle \quad\text{by
lemma~\ref{lem:heatcover}}.
\end{split}
\end{equation}
This equation makes it clear that the time-evolution of
$|\psi(\tau)\rangle$ is determined from that of $|\phi\rangle$ on
$X$.
\end{proof}

\section{Examples: the Hypercube $Q^n$ and Cayley and Schreier
Graphs}\label{sec:examples}

We now illustrate this result for the hypercube and the Cayley and
Schreier graphs of a finite group.

\begin{definition}
Let $G$ be a group, and let $S\subset G$ be a set of group
elements such that: (1) The identity element $e\not\in S$;  (2) If
$x\in S$, then $x^{-1}\in S$. The \emph{Cayley graph} $X(G,S)$
associated with $G$ and $S$ is then defined as the simple graph
having one vertex associated with each group element and directed
edges $(g, h)$ whenever $gh^{-1}\in S$. (It is easy to check that
with our definition of the generating set this graph is
well-defined.)
\end{definition}

\begin{example}
We think of the hypercube graph $Q_n$ of dimension $n$ as a Cayley
graph of the the abelian group $(\mathbb{Z}/2\mathbb{Z})^n$, the
$n$-fold direct product of $\mathbb{Z}/2\mathbb{Z}$. Specifically,
we write $Q_n =$  $ X((\mathbb{Z}/2\mathbb{Z})^n, \{e_1, e_2,
\ldots, e_n\})$, where $e_j = (0,\ldots,1,\ldots,0)$ is the unit
vector with a one in the $j$th entry. Note that $|V(Q_n)| = 2^n$.
It is easily verified that $Q_n$ is a covering for the weighted
path $P^Q_n$ of size $|V(P^Q_n)| = n+1$ with adjacency matrix
\begin{equation}\label{eq:qnadj}
A(P^Q_n) = \left[ \begin{matrix} 0 & \sqrt{n} & 0 &  \cdots & 0\\
\sqrt{n} & 0 & \sqrt{2(n-1)} &  \cdots & 0 \\
0 & \sqrt{2(n-1)} & 0 &
\cdots & 0 \\
\vdots &  &  & \ddots &  \sqrt{n}\\
0 & 0  & \cdots & \sqrt{n} & 0
\end{matrix}\right].
\end{equation}
The projection map $\pi:V(Q_n)\rightarrow V(P^Q_n)$ maps the
collection of vertices of $V(Q_n)$ of hamming weight $j$ to the
single vertex $v_j\in V(P^Q_n)$ at position $j$.
\end{example}

\begin{remark}
Because $Q_n$ is vertex-transitive (i.e.\ the automorphism group
$\aut(Q_n)$ acts transitively on $V(Q_n)$) a quantum walk on $Q_n$
beginning on the state concentrated on any single vertex is
equivalent to a walk on the weighted path $P^Q_n$ beginning at the
first vertex. (Note that all Cayley graphs $Y$ are
vertex-transitive, where the automorphism group $\aut(Y)$ acts
transitively on the vertex set.) The adjacency operator
(\ref{eq:qnadj}) of the weighted path is exactly the $x$ angular
momentum operator $J_x$ in the spin-$\frac{n}{2}$ irrep.\ of
$\text{{\sl SU}}(2)$. One can envisage the quantum state
propagating from the first vertex to the last vertex in $P^Q_n$ as
a localised state at $z=+\frac{n}{2}$ on north pole of the Bloch
sphere propagating across the sphere (and hence becoming
completely delocalised around the equator in the process) to the
south pole $z=-\frac{n}{2}$.
\end{remark}

Our second example concerns the Cayley graph for an arbitrary
finite abelian or nonabelian group $G$. To introduce it we need a
definition.

\begin{definition}
Given a subgroup $H$ of a finite group $G$ and a generating subset
$S$ (recall $s\in S$ implies $s^{-1}\in S$ and $e\not\in S$) of
$G$, which we call the edge set, the \emph{Schreier graph}
$X=X(G/H, S)$ has as vertices the cosets $gH$, $g\in G$. Two
vertices $gH$ and $s^{-1}gH$ are joined by an edge for all $s\in
S$.
\end{definition}

\begin{remark}
Note that unless $H$ is a normal subgroup $H\trianglelefteq G$
then the Schreier graph $X(G/H, S)$ will contain loops. In the
case that $H$ is normal then $X(G/H, S)$ is exactly the Cayley
graph of the group $G/H$ with generating set $S$. Note also that
our definition of graph covering spaces includes graphs with loops
(i.e.\ when $w(u,u)>0$) so that the following results are
well-defined.
\end{remark}

\begin{lemma}
The adjacency matrix $A(X)$ for a Cayley graph $X(G,S)$ can be
written in the following way
\begin{equation}
A(X) = \sum_{s\in S}\sum_{x\in G} |x\rangle\langle s^{-1}x|.
\end{equation}
Similarly, the adjacency matrix $A(X(G/H,S))$ for the Schreier
graph $X(G/H,S)$ is given by
\begin{equation}
A(X(G/H,S)) = \sum_{s\in S}\sum_{gH\in H} |gH\rangle\langle
s^{-1}gH|.
\end{equation}
\end{lemma}

We aim to show that if $Y$ is a Cayley graph $Y=X(G,S)$ for a
finite group $G$ with generating set $S$ then it is a covering
space for the Schreier graphs $X(G/H, S)$ for all $H\le G$. In
order to show this we define the projection map $\pi:Y\rightarrow
X(G/H, S)$ by $\pi(g) = gH$. The corresponding pull-back operator
is then given by
\begin{equation}
P = \sum_{\substack{ gH\in G/H
\\ g'\in gH }} \frac{1}{\sqrt{|H|}}|gH\rangle \langle g'|.
\end{equation}

In order to establish our claim we need to show that $PA(Y)$ and
$A(X(G/H,S))P$ are equal and apply lemma~\ref{lem:godsil1}:
\begin{equation}
\begin{split}
A(X(G/H,S))P &= \frac{1}{\sqrt{|H|}}\sum_{s\in S}\sum_{gH\in
G/H}\sum_{\substack{kH\in G/H\\ k'\in kH }} |gH\rangle\langle
s^{-1}gH|kH\rangle\langle k'| \\
&= \frac{1}{\sqrt{|H|}}\sum_{s\in S}\sum_{\substack{kH\in G/H\\
k'\in kH }}
\delta_{s^{-1}gH,kH}|gH\rangle\langle k'| \\
&= \frac{1}{\sqrt{|H|}}\sum_{s\in S}\sum_{\substack{kH\in G/H\\
k'\in kH }} |kH\rangle\langle s^{-1}k'|\\
&= PA(Y).
\end{split}
\end{equation}
The transition from the second-last line to the last follows by a
change of variable $kH\mapsto skH$ and $k'\mapsto sk'$, and the
definition of $P$ and $A(Y)$. This discussion, together with
lemma~\ref{lem:coverwalk}, constitutes the following corollary.

\begin{corollary}\label{cor:schreierwalk}
A quantum walk on a Cayley graph $X(G,S)$ which begins on the
fibre-constant state $|\phi_H\rangle = \sqrt{|H|/|G|}\sum_{gH\in
G/ H} c_{gH}|\pi^{-1}(gH)\rangle$ is isomorphic to a quantum walk
on induced the Schreier graph $X(G/H,S)$ for the subgroup $H$
which begins on the induced state $|\phi'\rangle
=\sqrt{|H|/|G|}\sum_{gH\in G/ H} c_{gH}|gH\rangle$.
\end{corollary}

\section{Efficient Gate Sequences for Continuous-Time Quantum
Walks}\label{sec:efficient}

In this section we explain how to construct efficient gate
sequences simulating continuous-time quantum walks on certain
classes of graphs $Y$. The principle feature intuitively exploited
in these gate sequences is that the graph $Y$ covers a tower
$Y\xrightarrow{\pi_{N}}X_{N}\xrightarrow{\pi_{N-1}}X_{N-1}
\xrightarrow{\pi_{N-2}}\cdots\xrightarrow{\pi_{1}}X_1$ of graphs
$X_j$. An interesting feature of these gate sequences is that
their length doesn't depend on the time $t$ the walk is simulated
for.

Consider the cycle $C_{2^n}$ on $2^n$ vertices, which is the
Cayley graph $C_{2^n} = X(\mathbb{Z}/2^n\mathbb{Z}, \{\pm 1\})$ of
the cyclic group. We begin by presenting a quantum circuit which
simulates the quantum walk on the cycle $C_{2^n}$ which uses only
$O(\poly(n))$ gates. A special feature of this gate sequence is
that it can simulate the walk for any given time $t$, including
exponential times $O(t) = 2^{\poly(n)}$.

Recall \cite{lubotzky:1995a} that the eigenvalues and eigenstates
of the adjacency matrix $A(C_{m})$ are given in terms of sums of
the characters $\mu^j = e^{i\frac{2\pi}{m}j}$, $j=0, \ldots, m-1$,
of $\mathbb{Z}/m\mathbb{Z}$. Specifically,
\begin{equation}
A(C_{m}) = \sum_{j=0}^{m-1} (\mu^j+\mu^{-j})|W(j)\rangle \langle
W(j)|,
\end{equation}
where
\begin{equation}
|W(j)\rangle = \frac{1}{\sqrt{m}}\sum_{k=0}^{m-1} \mu^{jk}
|k\rangle.
\end{equation}
Note that the vectors $|W(j)\rangle$ are precisely those given by
applying the quantum fourier transformation to $|j\rangle$:
\begin{equation}
|W(j)\rangle = \text{QFT}|j\rangle =
\frac{1}{\sqrt{N}}\sum_{k,l=0}^{m-1} \mu^{kl} |k\rangle\langle
l|j\rangle.
\end{equation}

Our objective is to simulate, with a quantum circuit, the
evolution
\begin{equation}
U[C_{2^n}](t) = \sum_{j=0}^{m-1} e^{-i 2\cos\left(\frac{2\pi}{m}
j\right) t} |W(j)\rangle\langle W(j)|,
\end{equation}
where $m=2^n$. We note that this unitary can be written as
\begin{equation}
U[C_{2^n}](t)=\text{QFT} \Phi(t) \text{QFT}^\dag,
\end{equation}
where
\begin{equation}
\Phi(t) = \sum_{j=0}^{m-1} e^{-i 2\cos\left(\frac{2\pi}{m}
j\right) t} |j\rangle\langle j|.
\end{equation}
The phase operation $\Phi(t)$ can be performed efficiently using,
for instance, the phase kickback trick \cite{cleve:1997a}. The
phase kickback trick shows that if the computation
\begin{equation}
|j\rangle \mapsto |j\rangle|f(j)\rangle,
\end{equation}
where $|j\rangle$ is a computational basis state of $n$ qubits and
$f(j)$ is an $n$ bit approximation to $(2\cos\left(\frac{2\pi}{m}
j\right) t) \mod 2\pi$ (calculable efficiently classically), is
implementable efficiently, then the phase changing operation
\begin{equation}
|j\rangle \mapsto e^{-i 2\cos\left(\frac{2\pi}{m} j\right) t}
|j\rangle,
\end{equation}
is implementable using $O(\poly(n))$ quantum gates
\cite{cleve:1997a}. (It is easy to obtain arbitrary accuracy by
appending more qubits to the register for $|f(x)\rangle$.)

Hence, because there is an efficient (i.e., using $O(\poly(n))$
elementary gates) quantum circuit for the quantum fourier
transform \cite{nielsen:2000a}, there is an efficient gate
sequence approximating the propagator $U[C_{2^n}](t)$. Moreover,
the number of gates required to simulate the propagator does not
depend on the time $t$.

We now note some features of our gate sequence for $U[C_m](t)$.
Firstly, we point out that the circuits for $U[C_m](t)$ can be
easily generalised to simulate quantum walks on a many other
families of graphs. This is because any circulant matrix is
diagonalised by the discrete fourier transformation, and so has
the same eigenvectors as $C_m$. This observation allows us to
generalise our simulation sequence to any circulant matrix whose
eigenvalue set $\{\lambda_j\}$ is efficiently calculable using
classical or quantum means. In particular, this means that all
Cayley graphs of $\mathbb{Z}/m\mathbb{Z}$ and the Paley
graphs\footnote{Given a finite field $\mathbb{F}_q$ with $q$
elements, the Paley graph $P(\mathbb{F}_q)$ is the graph with
vertex set $V=\mathbb{F}_q$ where two vertices are joined when
their difference is a square in the field. This is an undirected
graph when $q$ is congruent $1 (\text{mod} 4)$ \cite{chung:1997a}.
Note that Paley graphs have edge density approximately
$\frac{1}{2}$, and so are not efficiently simulatable using the
recipe of \cite{aharonov:2003a}.} are efficiently simulatable
using relatives of the quantum circuit we have constructed.
Additionally, any \emph{cartesian product} of graphs which are
efficiently simulatable via the gate sequence above will be
efficiently simulatable. This is because the adjacency matrix for
the cartesian product of two graphs $X$ and $Y$ is $A(X\times Y) =
A(X)\otimes I + I\otimes A(Y)$. In particular, a gate sequence for
the $m$-torus $T^m_{2^n}$ on $2^n$ vertices follows from
\begin{equation}
e^{-iA(T^m) t} = \bigotimes_{j=0}^{m-1} e^{-iA(C_{2^n})t}.
\end{equation}

The second feature we would like to point out is that the covering
space properties of $C_{2^n}$, in particular that $C_{2^n}$ is a
covering space for $C_{2^{n/2}}$, allow us to ``rediscover'',
using geometric constructions, an efficient quantum circuit for
the quantum fourier transform. Roughly speaking, an eigenvector of
the base graph $C_{2^{n/2}}$ can be lifted to $2^{n/2}$
eigenvectors of $C_{2^n}$ via a simple unitary recipe. This is a
general feature of the laplacian on a covering space, and extends
even to the continuous case \cite{rosenberg:1997a}. This idea is
explored further in \cite{osborne:2004a}.

We believe that the existence of covering spaces structures in
graphs may be utilised more generally to supply efficient quantum
circuits for quantum walks on graphs. We sketch one result which
is indicative of this idea.

We restrict our attention to regular graphs. Let $Y$ be a regular
graph on $2^n$ vertices which is a covering space for a tower of
graphs $Y\xrightarrow{\pi_{N}}X_{N}\xrightarrow{\pi_{N-1}}X_{N-1}
\xrightarrow{\pi_{N-2}}\cdots\xrightarrow{\pi_{1}}X_1$ with the
property that $|V(X_{j+1})|\approx \sqrt{|V(X_j)|}$. In this case
$N$ must be $O(\poly(\log(|V(Y|)))$.

We assume that $m=|\pi^{-1}_N(u)|$ is the same for all $u\in
V(X_N)$. Consider the \emph{transition matrix} $A(u,v)$ for an
edge $(u,v)\in E(X_N)$, which is the $2m\times 2m$ matrix whose
$(x,y)$ entry, where $x,y\in\pi^{-1}_N(u)\cup\pi^{-1}_N(v)$, is
given by $A(Y)_{x,y}$.

Suppose we can label the vertices in $\pi^{-1}_N(u)$ so that
\begin{equation}
A(u,v) = \left(\begin{matrix}0& 1\\ 1 &
0\end{matrix}\right)\otimes I_m,\quad \text{or} \quad A(u,v) =
\left(\begin{matrix}0& 1\\ 1 & 0\end{matrix}\right)\otimes
\frac{1}{\sqrt{m}}J_m,
\end{equation}
where $I_m$ is the $m\times m$ identity matrix and $J_m$ is the
$m\times m$ all $1$'s matrix. We call such a transition matrix
\emph{trivial}, and the we say that the edge $(u,v)$ has
\emph{trivial pull-back}.

Suppose the edges of $A(X_N)$ have trivial pull-back for all but
$O(\poly(\log(|V(Y)|)))$ edges. In the case we can write the
adjacency matrix for $A(Y)$ as
\begin{equation}\label{eq:adjdecomp}
A(Y) = A(X_N)\otimes I_m + \mathcal{D}, \quad \text{or} \quad A(Y)
= A(X_N)\otimes \frac{1}{\sqrt{l}}J_m + \mathcal{D},
\end{equation}
where $\mathcal{D}$ is a symmetric $|V(Y)|\times |V(Y)|$ matrix
whose $(x,y)$ entry, $x,y\in\pi^{-1}_N(u)\cup\pi^{-1}_N(v)$, is
nonzero only when there is an edge $(u,v)\in E(X_N)$ whose
transition matrix is nontrivial. When the edges of $A(X_N)$ have
trivial pull-back for all but $O(\poly(\log(|V(Y)|)))$ edges then
$\mathcal{D}$ is sparse, and has $O(\poly(\log(|V(Y)|)))$ nonzero
entries in each row (in the language of \cite{aharonov:2003a},
$\mathcal{D}$ is \emph{row sparse}).

We want to find a quantum circuit which simulates
$U[Y](t)=e^{-iA(Y) t}$ (note that because $Y$ is regular the
action of the adjacency matrix and laplacian are equivalent). To
do this we apply the Trotter formula (see \cite{nielsen:2000a} for
details and further discussion)
\begin{equation}
\lim_{n\rightarrow \infty} (e^{-iAt/n}e^{-iBt/n})^n = e^{-i
(A+B)t}.
\end{equation}
By taking $O(n)=\poly(\uinvnorm A \uinvnorm, \uinvnorm B\uinvnorm,
\log(|V(Y)|))$ we gain a good approximation to the time evolution
of $A+B$ for time $O(t) = \poly(n)$ \cite{nielsen:2000a} .

Applying the Trotter formula to (\ref{eq:adjdecomp}) we find that
\begin{equation}
U[Y](t) \approx ((e^{-iA(X_N)\otimes I_mt/n})
e^{-i\mathcal{D}t/n})^n, \quad \text{or} \quad U[Y](t) \approx
((e^{-iA(X_N)\otimes J_mt/n}) e^{-i\mathcal{D}t/n})^n.
\end{equation}
Because $\mathcal{D}$ is \emph{row sparse}, the simulation
algorithm of \cite{aharonov:2003a} can be applied to simulate
$e^{-i\mathcal{D}t/n}$ efficiently. (In order to guarantee the
applicability of the Trotter formula we assume that $\uinvnorm
A(Y) \uinvnorm$ grows polylogarithmically with $|V(Y)|$.)

We recursively reapply this construction to $A(X_N)$ (assuming, at
each step, that the pull-back of the edges $X_j$ is trivial for
all but a small number of edges.) until we have expressed all
instances of $A(Y)$ with $A(X_1)$. This construction furnishes an
efficient (i.e.\ using $O(\poly(\log(|V(Y)|)))$ elementary quantum
gates) quantum circuit which simulates the propagator $U[Y](t)$
accurately for times $t$ which are $O(\poly(\log(|V(Y)|)))$.

The calculations in this section should be seen as representative
of a general theory analogous to that initiated for quantum
fourier transforms in \cite{moore:2003a}. That is, given a tower
of covering spaces
$Y\xrightarrow{\pi_{N}}X_{N}\xrightarrow{\pi_{N-1}}X_{N-1}
\xrightarrow{\pi_{N-2}}\cdots\xrightarrow{\pi_{1}}X_1$, what are
the conditions the $X_j$ must satisfy in order to give rise to an
efficient gate sequence? We believe that such a theory is worth
developing because it will potentially provide a class of quantum
circuits whose behaviour may be interesting from an algorithmic
point of view. Interestingly, except for Cayley graphs, such
quantum circuits will be unrelated to discrete fourier transforms.

\section{The Hidden Cover Problem}\label{sec:hiddencover}
The discussion in the previous section indicates that
continuous-time quantum walks on certain graphs admit efficient
gate decompositions whose size doesn't depend on the length of
time the walk is simulated for. This feature can be exploited to
give polynomial time (in $\log(|V(Y)|)$) quantum algorithms which
can measure the hamiltonian $A(Y)$. As a simple corollary of this
we obtain an alternative observable for the hidden subgroup
problem which is not immediately equivalent to that employed by
Shor's algorithm (and related quantum algorithms). Motivated by
these new (efficiently implementable) observables we propose
another generalisation of the hidden subgroup problem: \emph{the
hidden covering space problem} {\sf HiddenCover}.

The (coherent sampling) hidden subgroup problem for a group $G$
(see, for example, \cite{nielsen:2000a} and references therein,
for a discussion of the hidden subgroup problem) consists of a
black box which outputs at random quantum states
$|\psi_{gH}\rangle$ which are equal superpositions of elements of
cosets of a hidden subgroup $H\le G$. (The Hilbert space
$\mathcal{H}$ in this case is taken to be the group algebra
$\mathbb{C}[G]$.) The problem is to determine $H$ using as few of
the coset states $|\psi_{gH}\rangle$ and as little quantum
computational time as possible.

The solution of the specific case $G=\mathbb{Z}/pq\mathbb{Z}$,
where $p$ and $q$ are prime and $H=\mathbb{Z}/p\mathbb{Z}$ or
$\mathbb{Z}/q\mathbb{Z}$, is well-known --- this is Shor's
factoring algorithm.

In the following we refer to a quantum state $|\phi\rangle$ in
$\mathbb{C}[\mathbb{Z}/n\mathbb{Z}]$, $n=pq$, as a
\emph{constant-coset} state on cosets of a subgroup
$\mathbb{Z}/q\mathbb{Z}\le \mathbb{Z}/pq\mathbb{Z}$ if it can be
written
\begin{equation}
|\phi\rangle = \sum_{j=0}^{p-1} c_j|\alpha_j\rangle,
\end{equation}
where $|\alpha_j\rangle = 1/\sqrt{q}\sum_{l=0}^q |j+lp\rangle$, $j
= 1,\ldots, p-1$.

The expansion of a constant coset state $|\phi\rangle$ in the
basis $|W(j)\rangle$ can be found via (this is essentially the
discrete Poisson summation formula \cite{terras:1999a}):
\begin{equation}\label{eq:expcoeffcoset}
\begin{split}
\langle W(k)|\alpha_j\rangle &= \frac{1}{q\sqrt{p}}\sum_{l=0}^q
e^{-\frac{2\pi i}{pq}(j+lp)k} \\
&= \frac{e^{-\frac{2\pi i}{pq}jk}}{q\sqrt{p}}\sum_{l=0}^q
e^{-\frac{2\pi i}{q} lk} \\
&= \frac{e^{-\frac{2\pi i}{pq}jk}}{\sqrt{p}}\delta_{k,\lambda q},
\quad \lambda = 0,\ldots,p-1.
\end{split}
\end{equation}
The expansion coefficients $\langle W(k)|\alpha_j\rangle$ are
nonzero only when $k$ is a multiple of $q$.

In the abelian hidden subgroup problem we have a black box which
outputs at random the constant-coset states $|\alpha_j\rangle$.
The standard solution of the hidden subgroup problem proceeds by
applying the quantum fourier transform to the $|\alpha_j\rangle$
and then measuring in the computational basis. This yields an
approximation $\widetilde{r/q}$ to the number $r/q$, for random
integer $r$. After enough samples the identity of the hidden
subgroup can be inferred by applying the continued-fractions
algorithm to $\widetilde{r/q}$.

We now supply an alternative procedure to the standard quantum
fourier transform which uses a quantum walk on the Cayley graph
$X(\mathbb{Z}/pq\mathbb{Z},\{\pm 1\})$. We don't claim that this
is any different to the standard quantum fourier transform
algorithm for the HSP on $\mathbb{Z}/pq\mathbb{Z}$. The point is
that the generalisation of this procedure to other graphs will
\emph{not} be equivalent to the quantum fourier transform method.
At the moment, however, we can only perform the algorithm for
cyclic groups.

The eigenvalues of the hamiltonian (i.e.\ the Laplacian
$\triangle$) for the quantum walk on the cycle
$X(\mathbb{Z}/pq\mathbb{Z}, \{\pm 1\})$ are given by
\begin{equation}
\lambda_j = \cos\left(\frac{2\pi}{pq}j\right).
\end{equation}
Consequently, if the hamiltonian $\triangle [X]$ is measured
exactly on a constant-coset state $|\phi\rangle$ then, by the
discussion surrounding (\ref{eq:expcoeffcoset}), the only
eigenvalues that can be measured are those of the form $\lambda_{j
q}=\cos\left(\frac{2\pi}{p}j\right)$, for some random $0\le j\le
q-1$.

We can effectively measure $\triangle[X]$ using $U[X](t)$ ---
which, as discussed in \S\ref{sec:efficient}, can be implemented
efficiently, using the discretisation of von Neumann's
prescription for measuring a hermitian operator given by
\cite{childs:2002b}. Implementing this measurement yields an
approximation $\widetilde{\lambda}_j$ to an eigenvalue of
$\triangle[X]$ for random $j$. Because $\cos(x)$ is continuous,
the function
\begin{equation}
\widetilde{j/q} = \cos^{-1}(\widetilde{\lambda}_j)
\end{equation}
is a good approximation to the ratio ${j/q}$, for random $0\le
j\le q-1$. Applying the continued fractions algorithm yields $q$.

The previous result is suggestive of generalisations in the
following way.

Imagine we have a graph $Y$ which is a covering space
$\pi:Y\rightarrow X$ for $X$. Imagine, further, we have a black
box which outputs at random fibre-constant states
$|\psi_{\pi}\rangle$. Recall that any fibre-constant state can be
written in terms of the pull-backs of the eigenstates of $X$:
\begin{equation}
|\psi_{\pi}\rangle = \sum_{j=0}^{|V(X)|-1} c_j P^\dag
|E_j(X)\rangle,
\end{equation}
where $|E_j(X)\rangle$ are the eigenstates of $X$.

\emph{This means that if the hamiltonian $\triangle[Y]$ is
measured on $|\psi_{\pi}\rangle$ then it can only report
eigenvalues of $X$, rather than eigenvalues from the full possible
spectrum of $Y$. If the spectrum of $\lambda(X)$ can be
distinguished from the spectra $\lambda(Z_l)$ of the other graphs
$Z_l$ that $Y$ is a covering space for then this measurement can
identify the hidden graph $X$. As we showed previously, in the
case where $Y=X(\mathbb{Z}/pq\mathbb{Z}, \{\pm 1\})$, and
$X=X(\mathbb{Z}/q\mathbb{Z}, \{\pm 1\})$, then this is equivalent
to solving the abelian hidden subgroup problem.}

Thus, generalising boldly, we believe that quantum computers ought
to be able to efficiently solve the following problem.

\begin{framed}
{\large \bf The hidden covering space problem {\sf HiddenCover}}
\\
\noindent {\bf Input:}
\begin{enumerate}
\item A class $\mathcal{C}$ of mathematical objects.%
\item A quantisation scheme which associates a unitary matrix
$U(Y)$, with each $Y\in \mathcal{C}$, acting on an associated
Hilbert space $\mathcal{H}(Y)$. %
\item An object $Y$ from $\mathcal{C}$ with the promise that $Y$
is a covering space $\pi_l:Y\rightarrow X_l$ only for a (known)
set of objects $X_l$, $l=0, \ldots, n$. %
\item A black box which randomly emits fibre-constant states
$|\psi_{\pi_l}\rangle$ for some (unknown) projection $\pi_l$, for
some $l=0, \ldots, n$.
\end{enumerate}
\noindent {\bf Task:} Determine the projection $\pi_l$, and hence
identify the base space $X_l$.
\end{framed}

We are willing to conjecture that {\sf HiddenCover} is efficiently
solvable on a quantum computer for certain classes of simple
graphs using continuous-time quantum walks.

A natural question arises: what happens when we apply the
algorithms sketched above to the Cayley graph for the dihedral
group $D_n = \langle s,t \,|\, s^n = t^2 = e\rangle$? In this
case, we are after a hidden transposition $\langle e,
s^jt\rangle$. Unfortunately, it can be verified that the Schreier
graphs for the $n$ hidden transpositions are isospectral, so that,
with one fibre-constant state $|\psi_{\pi}\rangle$ it is
impossible to tell which transposition $\langle e, s^jt\rangle$ is
hidden.

\section{Conclusions and Future Directions}
In this paper we have explored two related ideas, both of which
are connected with covering space structures. The first is that a
quantum evolution for an object $Y$ which is a covering space for
another object $X$ can be equivalent to an evolution on the
smaller object $X$. The second is that covering space structures
can be exploited to give efficient gate sequences for their
quantisations. The first idea can be exploited to give
hitting-time speedups for such evolutions as quantum walks.
Additionally, the first idea leads to the notion that hidden
covering spaces can be identified spectrally. The second leads to
the notion that quantum computers can do this efficiently.

We shall conclude with a list of future directions.
\begin{enumerate}
\item For the dihedral group consider walking on a graph $Y$ which
is not a simple cartesian product of $\log|D_n|$ copies of the
Cayley graph of $D_n$ (this is what happens when you measure the
propagator $\log|D_n|$ times). Possible graphs to try might be
certain graph products of the Cayley graph $X(D_n, S)$ for the dihedral group with generating set $S$. %
\item What about other mathematical objects? There are some
promising candidates, such as algebraic number fields, smooth
manifolds, and knots which have natural structures amenable to
quantisation. %
\item What sorts of computational problems are expressible as
variants of {\sf HiddenCover}? Are there any interesting
computational problems?
\end{enumerate}

Finally, a word on discrete-time quantum walks. Discrete-time
quantum walks represent another quantisation scheme for simple
graphs where the topology of the graph is encoded in the unitary
quantisation. It is natural to ask how the ideas of this paper
extend to the discrete-time quantum walks? It is interesting to
remark that a coined quantum walk on a graph $X$ is in fact a
discrete-time quantum walk on a certain directed graph $Y$ (namely
$Y$ is the line digraph of $X$ --- see, for example,
\cite{gusfield:1998a} for an application of line digraph in the
design of algorithms) which is homomorphic to $X$. It can be
observed that $Y$ covers $X$ (the spectrum of $Y$ is the spectrum
of $X$ plus a zero eigenvalues with the appropriate multiplicity
\cite{rosenfeld:2001a}). Now, the construction mentioned above is
only one possible quantisation of graphs. Are there other natural
quantisations? Perhaps one can pick out a canonical quantisation
by demanding that it respect covering space structures?

Given a matrix $M$ (over any field), a directed graph $X$ is said
to be the \emph{graph of} $M$ if the $uv$-th entry of $M$ is
nonzero if and only if there is a directed arc $(u,v)$ for every
pair of vertices $u,v$. A characterisation of graphs of unitary
matrices is still missing (see, for example, the open problems
section of \cite{zyczkowski:2003a}). In general, what is the
relation between graphs of unitary matrices and covering spaces?
The study of this problem may be useful in understanding the
combinatorial properties of certain combinatorial designs like
weighing matrices.

\subsection*{Acknowledgments} We are very grateful to Andreas Winter
for many helpful and inspiring discussions as well as for many
helpful comments on this paper. We are also very grateful to
Andrew Childs for carefully reading this paper and many helpful
suggestions. TJO is grateful to the EU for support for this
research under the IST project RESQ. SS was supported by a
University of Bristol Research Scholarship.

\newcommand{\etalchar}[1]{$^{#1}$}
\providecommand{\bysame}{\leavevmode\hbox
to3em{\hrulefill}\thinspace}


\begin{thebibliography}{AvDK{\etalchar{+}}04}
\providecommand{\MR}{\relax\ifhmode\unskip\space\fi MR }
\providecommand{\MRhref}[2]{%
  \href{http://www.ams.org/mathscinet-getitem?mr=#1}{#2}
} \providecommand{\href}[2]{#2} \expandafter\ifx\csname
url\endcsname\relax
  \def\url#1{\texttt{#1}}\fi
\expandafter\ifx\csname
urlprefix\endcsname\relax\def\urlprefix{URL}\fi
\providecommand{\eprint}[2][]{\url{#2}}

\bibitem[AKR04]{ambainis:2004a}
Andris Ambainis, Julia Kempe, and Alexander Rivosh, \emph{Coins
{M}ake
  {Q}uantum {W}alks {F}aster}, 2004, \eprint{quant-ph/0402107}

\bibitem[Amb03]{ambainis:2003a}
Andris Ambainis, \emph{Quantum walk algorithm for element
distinctness}, 2003,
  \eprint{quant-ph/0311001}

\bibitem[ATS03]{aharonov:2003a}
Dorit Aharonov and Amnon Ta-Shma, \emph{Adiabatic quantum state
generation and
  statistical zero knowledge}, Proceedings of the 35th Annual ACM Symposium on
  Theory of Computing held in San Diego, CA, June 9--11, 2003 (New York),
  Association for Computing Machinery (ACM), 2003, pp.~20--29,
  \eprint{quant-ph/0301023}

\bibitem[AvDK{\etalchar{+}}04]{aharonov:2004a}
D.~Aharonov, W.~van Dam, J.~Kempe, Z.~Landau, S.~Lloyd, and
O.~Regev, \emph{On
  the universality of adiabatic quantum computation}, 2004

\bibitem[Big93]{biggs:1993a}
Norman Biggs, \emph{Algebraic graph theory}, 2nd ed., Cambridge
Mathematical
  Library, Cambridge University Press, Cambridge, 1993; \MR{95h:05105}

\bibitem[BV97]{bernstein:1997a}
Ethan Bernstein and Umesh Vazirani, \emph{Quantum complexity
theory}, SIAM J.
  Comput. \textbf{26} (1997), no.~5, 1411--1473; \MR{99a:68053}

\bibitem[CCD{\etalchar{+}}03]{childs:2003a}
Andrew~M. Childs, Richard Cleve, Enrico Deotto, Edward Farhi, Sam
Gutmann, and
  Daniel~A. Spielman, \emph{Exponential algorithmic speedup by a quantum walk},
  Proceedings of the 35th Annual ACM Symposium on Theory of Computing held in
  San Diego, CA, June 9--11, 2003 (New York), Association for Computing
  Machinery (ACM), 2003, pp.~59--68

\bibitem[CDF{\etalchar{+}}02]{childs:2002b}
Andrew~M. Childs, Enrico Deotto, Edward Farhi, Jeffrey Goldstone,
Sam Gutmann,
  and Andrew~J. Landahl, \emph{Quantum search by measurement}, Phys. Rev. A
  \textbf{66} (2002), no.~3, 032314, \eprint{quant-ph/0204013}

\bibitem[CDS95]{cvetkovic:1995a}
Drago{\v{s}}~M. Cvetkovi{\'c}, Michael Doob, and Horst Sachs,
\emph{Spectra of
  graphs}, 3rd ed., Johann Ambrosius Barth, Heidelberg, 1995; \MR{96b:05108}

\bibitem[CE03]{childs:2003b}
Andrew~M. Childs and Jason~M. Eisenberg, \emph{Quantum algorithms
for subset
  finding}, 2003, \eprint{quant-ph/0311038}

\bibitem[CEMM98]{cleve:1997a}
R.~Cleve, A.~Ekert, C.~Macchiavello, and M.~Mosca, \emph{Quantum
algorithms
  revisited}, R. Soc. Lond. Proc. Ser. A Math. Phys. Eng. Sci. \textbf{454}
  (1998), no.~1969, 339--354, Quantum coherence and decoherence (Santa Barbara,
  CA, 1996), \eprint{quant-ph/9708016}; \MR{99m:81030}

\bibitem[CFG02]{childs:2002a}
Andrew~M. Childs, Edward Farhi, and Sam Gutmann, \emph{An example
of the
  difference between quantum and classical random walks}, Quantum Inf. Process.
  \textbf{1} (2002), no.~1-2, 35--43, \eprint{quant-ph/0103020}; \MR{1 964 317}

\bibitem[CG03]{childs:2003c}
Andrew~M. Childs and Jeffrey Goldstone, \emph{Spatial seach by
quantum walk},
  2003, \eprint{quant-ph/0306054}

\bibitem[Chu97]{chung:1997a}
Fan R.~K. Chung, \emph{Spectral graph theory}, CBMS Regional
Conference Series
  in Mathematics, vol.~92, Published for the Conference Board of the
  Mathematical Sciences, Washington, DC, 1997; \MR{97k:58183}

\bibitem[CY99]{chung:1999a}
Fan Chung and S.-T. Yau, \emph{Coverings, heat kernels and
spanning trees},
  Electron. J. Combin. \textbf{6} (1999), no.~1, Research Paper 12, 21 pp.\
  (electronic); \MR{2000g:35079}

\bibitem[GKWS98]{gusfield:1998a}
Dan Gusfield, Richard Karp, Lusheng Wang, and Paul Stelling,
\emph{Graph
  traversals, genes and matroids: an efficient case of the travelling salesman
  problem}, Discrete Appl. Math. \textbf{88} (1998), no.~1-3, 167--180;
  \MR{99k:90133}

\bibitem[GM80]{godsil:1980a}
C.~D. Godsil and B.~D. McKay, \emph{Feasibility conditions for the
existence of
  walk-regular graphs}, Linear Algebra Appl. \textbf{30} (1980), 51--61;
  \MR{83d:05068}

\bibitem[Hat02]{hatcher:2002a}
Allen Hatcher, \emph{Algebraic topology}, Cambridge University
Press,
  Cambridge, 2002; \MR{2002k:55001}

\bibitem[Kem03a]{kempe:2003b}
Julia Kempe, \emph{Quantum random walks: an introductory
overview}, Contemp.
  Phys. \textbf{44} (2003), no.~4, 307--327, \eprint{quant-ph/0303081}

\bibitem[Kem03b]{kempe:2003a}
Julia Kempe, \emph{Quantum {R}andom {W}alks {H}it {E}xponentially
{F}aster},
  RANDOM-APPROX (Sanjeev Arora, Klaus Jansen, Jos{\'e} D.~P. Rolim, and Amit
  Sahai, eds.), Lecture Notes in Computer Science, vol. 2764, Springer, 2003,
  pp.~354--369, \eprint{quant-ph/0205083}

\bibitem[Lub95]{lubotzky:1995a}
Alexander Lubotzky, \emph{Cayley graphs: eigenvalues, expanders
and random
  walks}, Surveys in combinatorics, 1995 (Stirling), London Math. Soc. Lecture
  Note Ser., vol. 218, Cambridge Univ. Press, Cambridge, 1995, pp.~155--189;
  \MR{96k:05081}

\bibitem[ML98]{maclane:1998a}
Saunders Mac~Lane, \emph{Categories for the working
mathematician}, 2nd ed.,
  Graduate Texts in Mathematics, vol.~5, Springer-Verlag, New York, 1998;
  \MR{2001j:18001}

\bibitem[MR02]{moore:2002a}
Cristopher Moore and Alexander Russell, \emph{Quantum walks on the
hypercube},
  RANDOM (Jos{\'e} D.~P. Rolim and Salil~P. Vadhan, eds.), Lecture Notes in
  Computer Science, vol. 2483, Springer, 2002, pp.~164--178,
  \eprint{quant-ph/0104137}

\bibitem[MRR03]{moore:2003a}
Cristopher Moore, Daniel Rockmore, and Alexander Russell,
\emph{Generic
  {Q}uantum {F}ourier {T}ransforms}, To appear in SODA 2004,
  \eprint{quant-ph/0304064}

\bibitem[MSS03]{magniez:2003a}
Fr{\'e}d{\'e}ric Magniez, Miklos Santha, and Mario Szegedy,
\emph{An
  ${O}(n^{1.3})$ {Q}uantum {A}lgorithm for the {T}riangle {P}roblem}, 2003,
  \eprint{quant-ph/0310134}

\bibitem[NC00]{nielsen:2000a}
Michael~A. Nielsen and Isaac~L. Chuang, \emph{Quantum computation
and quantum
  information}, Cambridge University Press, Cambridge, 2000; \MR{1 796 805}

\bibitem[Osb04]{osborne:2004a}
Tobias~J. Osborne, \emph{Efficient {Q}uantum {G}ate {S}equences
from {G}raph
  {C}overing {S}paces}, 2004

\bibitem[Pre98]{preskillnotes}
John Preskill, {P}hysics 229: {A}dvanced {M}athematical {M}ethods
of {P}hysics
  --- {Q}uan\-tum {C}omputation and {I}nformation. {C}alifornia {I}nstitute of
  {T}echnology, \texttt{http://www.theory.caltech/edu/people/}
  \texttt{preskill/ph229/}, 1998

\bibitem[RB01]{briegel:2001b}
Robert Raussendorf and Hans~J. Briegel, \emph{A {O}ne-{W}ay
{Q}uantum
  {C}omputer}, Phys. Rev. Lett. \textbf{86} (2001), no.~22, 5188--5191,
  \eprint{quant-ph/0010033}

\bibitem[Ros97]{rosenberg:1997a}
Steven Rosenberg, \emph{The {L}aplacian on a {R}iemannian
manifold}, London
  Mathematical Society Student Texts, vol.~31, Cambridge University Press,
  Cambridge, 1997, An introduction to analysis on manifolds; \MR{98k:58206}

\bibitem[Ros01]{rosenfeld:2001a}
Vladimir~Raphael Rosenfeld, \emph{Some spectral properties of the
arc-graph},
  Match (2001), no.~43, 41--48; \MR{2002c:05115}

\bibitem[Sho94]{shor:1994a}
Peter~W. Shor, \emph{Algorithms for quantum computation: discrete
logarithms
  and factoring}, 35th Annual Symposium on Foundations of Computer Science
  (Santa Fe, NM, 1994), IEEE Comput. Soc. Press, Los Alamitos, CA, 1994,
  pp.~124--134, \eprint{quant-ph/9508027}; \MR{1 489 242}

\bibitem[SKW03]{shenvi:2003a}
Neil Shenvi, Julia Kempe, and K.~Birgitta Whaley, \emph{Quantum
random-walk
  search algorithm}, Phys. Rev. A \textbf{67} (2003), no.~5, 052307,
  \eprint{quant-ph/0210064}

\bibitem[ST96]{stark:1996a}
H.~M. Stark and A.~A. Terras, \emph{Zeta functions of finite
graphs and
  coverings}, Adv. Math. \textbf{121} (1996), no.~1, 124--165; \MR{98b:11094}

\bibitem[ST00]{stark:2000a}
H.~M. Stark and A.~A. Terras, \emph{Zeta functions of finite
graphs and
  coverings. {II}}, Adv. Math. \textbf{154} (2000), no.~1, 132--195;
  \MR{2002f:11123}

\bibitem[ST01]{stark:1999a}
H.~M. Stark and A.~A. Terras, \emph{Artin {$L$}-functions of graph
coverings},
  Dynamical, spectral, and arithmetic zeta functions (San Antonio, TX, 1999),
  Contemp. Math., vol. 290, Amer. Math. Soc., Providence, RI, 2001,
  pp.~181--195; \MR{2003c:11113}

\bibitem[Sze04]{szegedy:2004b}
Mario Szegedy, \emph{Spectra of {Q}uantized {W}alks and a
  $\sqrt{\delta\epsilon}$-{R}ule}, 2004, \eprint{quant-ph/0401053}

\bibitem[Ter99]{terras:1999a}
Audrey Terras, \emph{Fourier analysis on finite groups and
applications},
  London Mathematical Society Student Texts, vol.~43, Cambridge University
  Press, Cambridge, 1999; \MR{2000d:11003}

\bibitem[Ter02]{terras:2002a}
Audrey Terras, \emph{Finite quantum chaos}, Amer. Math. Monthly
\textbf{109}
  (2002), no.~2, 121--139; \MR{1 903 150}

\bibitem[{\.Z}KSS03]{zyczkowski:2003a}
Karol {\.Z}yczkowski, Marek Ku{\'s}, Wojciech S{\l}omczy{\'n}ski,
and
  Hans-J{\"u}rgen Sommers, \emph{Random unistochastic matrices}, J. Phys. A
  \textbf{36} (2003), no.~12, 3425--3450; \MR{1 986 428}

\end{thebibliography}
\end{document}